\documentclass[12pt]{article}
\usepackage{graphicx}

\begin{document}

\title{Effects of polarization in \\electromagnetic processes 
\\in oriented crystals at high energy}

\author{V. N. Baier and V. M. Katkov\\
Budker Institute of Nuclear Physics\\ 630090 Novosibirsk, Russia}

\maketitle

\begin{abstract}
 
Under quite generic assumptions the general expression 
is derived for the probability of circularly polarized photon 
emission from the longitudinally polarized electron and for the
probability of pair creation of longitudinally polarized electron
(positron) by circularly polarized photon in oriented crystal
in a frame of the quasiclassical operator method.
For small angle of incidence the expression turns into
constant field limit with corrections due to inhomogeneous 
character of field in crystal. For relatively large angle 
of incidence the expression gets over into the theory of 
coherent radiation or pair creation. It is shown that the crystal
is a very effective device for helicity transfer 
from an electron to photon and back from a photon to 
electron or positron.

\end{abstract}


\section{Introduction}

The study of processes with participation of polarized electrons 
and photons permits to obtain important physical information.
Because of this experiments with use of polarized particles
are performed and are planning in many laboratories (BINP, CERN, 
SLAC, Jefferson Natl Accl Fac, etc). In this paper it is 
shown that oriented crystal is a unique tool for work with polarized
electrons and photons
 
The quasiclassical operator method developed by authors
\cite{BK0}-\cite{BK2} is adequate for consideration of the
electromagnetic processes at high energy. 
The probability of photon emission has a form (see
\cite{BKS}, p.63, Eq.(2.27); the method is given also in
\cite{BLP},\cite{BKF})
\begin{equation}
dw=\frac{e^2}{(2\pi)^2} \frac{d^3k}{\omega}
\int_{}^{}dt_2\int_{}^{}dt_1 R^{\ast}(t_2) R(t_1)
\exp \left[-\frac{i\varepsilon}{\varepsilon'}
\left(kx_2-kx_1\right) \right],
\label{1}
\end{equation}
where $k^{\mu}=(\omega, {\bf k})$ is the 4-momentum of the emitted
photon, $k^2=0$, $x^{\mu}(t)=(t, {\bf r}(t)),
~x_{1,2} \equiv x(t_{1,2}),~t$ is the time,
and ${\bf r}(t)$ is the particle location on a classical trajectory,
$kx(t)=\omega t- {\bf kr}(t),~\varepsilon$ is the energy of initial
electron, $\varepsilon'=\varepsilon-\omega$, 
we employ units such that $\hbar=c=1$.
The matrix element $R(t)$ is defined by the structure of a current.
For an electron (spin 1/2 particle) one has
\begin{eqnarray}
&& R(t)=\frac{m}{\sqrt{\varepsilon \varepsilon'}}
\overline{u}_{s_{f}}({\bf p'})\hat{e}^{\ast}u_{s_{i}}({\bf p})
=\varphi_{\zeta_f}^{+}\left(A(t)+i\mbox{\boldmath$\sigma$}{\bf B}(t) \right)
\varphi_{\zeta_i}, 
\nonumber \\
&& A(t)= \frac{1}{2}\left(1+\frac{\varepsilon}{\varepsilon'} \right)
{\bf e}^{\ast}\mbox{\boldmath$\vartheta$}(t), 
\nonumber \\
&& {\bf B(t)}=\frac{\omega}{2\varepsilon'}\left({\bf e}^{\ast} \times
\left(\frac{{\bf n}}{\gamma}- \mbox{\boldmath$\vartheta$}(t)\right) \right),
\label{2}
\end{eqnarray}
here  ${\bf e}$ is the vector of the polarization of a photon
(the Coulomb gauge is used),the four-component spinors $u_{s_f},
u_{s_i}$ describe the initial ($s_i$) and final ($s_f$)
polarization of the electron 
and we use for the description of electron polarization 
the vector  $\mbox{\boldmath$\zeta$}$  describing the
polarization of the electron (in its rest frame),
$\mbox{\boldmath$\zeta$}_i$ is the spin vector of initial electron,
$\mbox{\boldmath$\zeta$}_f$ is the spin vector of final electron,
the two-component spinors $\varphi_{\zeta_i},
\varphi_{\zeta_f}$ describe the initial and final 
polarization of the electron, {\bf v}={\bf v}(t) is the electron
velocity, $\mbox{\boldmath$\vartheta$}(t) = \left({\bf v}-{\bf
n}\right) \simeq {\bf v}_{\perp}(t)$, ${\bf
v}_{\perp}$ is the component of particle velocity perpendicular
to the vector ${\bf n}={\bf k}/|{\bf k}|$, $\gamma=\varepsilon/m$
is the Lorentz factor. The expressions in Eq.(\ref{2}) are
given for radiation of ultrarelativistic electrons, they are
written down with relativistic accuracy (terms $\sim 1/\gamma$
are neglected) and in the small angle approximation.

The important parameter $\chi$
characterizes the quantum effects in an external field, when
$\chi \ll 1$ we are in the classical domain and with $\chi \geq
1$ we are already well inside the quantum domain while for pair 
creation the corresponding parameter is $\kappa$ 
\begin{equation}
\chi =\frac{|{\bf F}|\varepsilon}{F_0m},\quad 
\kappa =\frac{|{\bf F}|\omega}{F_0m}, \quad {\bf F}={\bf
E}_{\perp}+({\bf v}\times {\bf H}),\quad {\bf E}_{\perp}={\bf E}-
{\bf v}({\bf v}{\bf E}),
\label{3}
\end{equation}
where ${\bf E} ({\bf H})$ is an electric (magnetic) field,
$F_0=m^2/e=(m^2c^2/e\hbar)$ is the quantum boundary (Schwinger)
field: $H_0=4.41 \cdot 10^{13} {\rm Oe},~ E_0=1.32\cdot 10^{16} 
{\rm V/cm}$. 

The quasiclassical operator method is applicable when $H \ll H_0,~
E \ll E_0$ and $\gamma \gg 1$.

Summing the combination $R^{\ast}(t_2) R(t_1)=R_2^{\ast}R_1$
over final spin states we have
\begin{equation}
\sum_{\zeta_f}R_2^{\ast}R_1 =A_2^{\ast}A_1
+\textbf{B}_2^{\ast}\textbf{B}_1
+i\left[A_2^{\ast}(\mbox{\boldmath$\zeta$}_i\textbf{B}_1)
-A_1  (\mbox{\boldmath$\zeta$}_i\textbf{B}_2^{\ast})
+\mbox{\boldmath$\zeta$}_i\left(\textbf{B}_2^{\ast}\times
\textbf{B}_1\right) \right],  
\label{4} 
\end{equation} 
where the two first terms describe the radiation of unpolarized 
electrons and the last terms is an addition dependent on 
the initial spin.

For the longitudinally polarized initial electron and for circular 
polarization of emitted photon we have \cite{BK4}
\begin{equation}
\sum_{\zeta_f} R_2^{\ast}R_1 = \frac{1}{4\varepsilon'^2}
\left\{\frac{\omega^2}{\gamma^2}(1+\xi)+
\left[(1+\xi)\varepsilon^2+(1-\xi)\varepsilon'^2
\right]\mbox{\boldmath$\vartheta$}_1\mbox{\boldmath$\vartheta$}_2 \right\},
\label{5}
\end{equation}
where $\xi=\lambda\zeta,~\lambda=\pm 1$ is the helicity of emitted photon,
$\zeta =\pm 1$ is the helicity of the initial electron. 
In this expression we omit the terms which vanish after integration
over angles of emitted photon.

The probability of
polarized pair creation by a circularly polarized photon
can be found from Eqs.(\ref{1}), (\ref{2}) using standard substitutions:
\begin{eqnarray}
&& \varepsilon' \rightarrow \varepsilon',\quad
\varepsilon \rightarrow -\varepsilon,\quad
\omega \rightarrow -\omega,\quad
\lambda \rightarrow  -\lambda, 
\nonumber \\
&& \zeta_i \equiv \zeta \rightarrow -\zeta',\quad
\zeta_f \equiv \zeta' \rightarrow \zeta,\quad
\xi \rightarrow \xi,\quad
\xi' \rightarrow -\xi'.
\label{6}
\end{eqnarray}  

Performing the substitutions Eq.(\ref{6}) in Eq.(\ref{5})
the combination for creation of pair with polarized positron 
by circularly polarized photon
\begin{equation}
\sum_{\xi'}R_{2}
R_{1}^{\ast}
=\frac{m^2}{8\varepsilon^2\varepsilon'^2}
\left\{\omega^2(1+\xi)
+\gamma^2\mbox{\boldmath$\vartheta$}_2\mbox{\boldmath$\vartheta$}_1
\left[\varepsilon^2(1+\xi) 
+\varepsilon'^2(1-\xi)\right]\right\}. 
\label{7}
\end{equation} 

It should be noted that a few different spin correlations are
known in an external field. But after averaging over directions of
crystal field only the longitudinal polarization considered here 
survives.

\section{General approach to electromagnetic processes in oriented crystals}

The theory of high-energy electron radiation and electron-positron pair
creation in oriented crystals was developed in \cite{BKS1}-\cite{BKS2},
and given in \cite{BKS}. In these publications the radiation
from unpolarized electrons was considered including 
the polarization density matrix of emitted photons. 
Since Eqs.(\ref{5}), (\ref{7}) have the same structure 
as for unpolarized particles, below we use systematically 
the methods of mentioned papers
to obtain the characteristics of radiation and pair creation
for longitudinally polarized particles.

Let us remind that along with the parameter $\chi$ which 
characterizes the quantum
properties of radiation there is another parameter
\begin{equation}
\varrho=2\gamma^2\left\langle (\Delta\textbf{v} )^2\right\rangle,
\label{1.1}
\end{equation} 
where $\left\langle (\Delta\textbf{v} )^2\right\rangle=
\left\langle \textbf{v}^2 \right\rangle -\left\langle \textbf{v} \right\rangle^2$
and $\left\langle\ldots \right\rangle$ denotes averaging over time.
In the case $\varrho \ll 1$ the radiation is of a dipole nature and it is formed
during the time of the order of the period of motion. In the case 
$\varrho \gg 1$ the radiation is of magnetic bremsstrahlung nature and
it is emitted from a small part of the trajectory.

In a crystal the parameter $\varrho$ depends on the angle of 
incidence $\vartheta_0$
which is the angle between an axis (a plane) of crystal and the momentum
of a particle. If $\vartheta_0 \leq \vartheta_c$ (where $\vartheta_c \equiv
(2V_0/\varepsilon)^{1/2},~V_0$ is the scale of continuous potential of 
an axis or a plane relative to which the angle $\vartheta_0$ is defined)
electrons falling on a crystal are captured into channels or low above-barrier
states, whereas for $\vartheta_0 \gg \vartheta_c$ the incident particles 
move high above the barrier. In later case we can describe the motion using 
the approximation of the  rectilinear trajectory, 
for which we find from Eq.(\ref{1.1}) the
following estimate 
$\varrho(\vartheta_0)=\left(2V_0/m\vartheta_0 \right)^2$.
For angles of incidence in the range $\vartheta_0 \leq \vartheta_c$
the transverse (relative to an axis or a plane) velocity of particle
is $v_{\perp}\sim \vartheta_c$ and the parameter obeys $\varrho \sim \varrho_c$
where $\varrho_c=2V_0\varepsilon/m^2$.
This means that side by side with the Lindhard angle $\vartheta_c$ the
problem under consideration has another characteristic angle
$\vartheta_v=V_0/m$ and $\varrho_c=(2\vartheta_v/\vartheta_c)^2$.

We consider here the photon emission (or pair creation) 
in a thin crystal  when the condition 
$\varrho_c \gg 1$ is satisfied.
In this case the extremely difficult task of averaging of 
Eqs.(\ref{5}),(\ref{7}),  derived for a given trajectory,  over all 
possible trajectories of electrons in a crystal  simplifies radically.
In fact, if $\varrho_c \gg 1$ then in the range where trajectories are
essentially non-rectilinear ($\vartheta_0 \leq \vartheta_c,~v_{\perp} \sim 
\vartheta_c$) the mechanism of photon emission is of the magnetic 
bremsstrahlung nature and the characteristics of radiation can be expressed
in terms of local parameters of motion. Then the averaging procedure can be  
carried out simply if one knows the distribution function in the transverse
phase space $dN(\varrho, \textbf{v}_{\perp})$, which for a thin crystal 
is defined directly by the initial conditions of incidence of particle
on a crystal.

Substituting Eq.(\ref{5}) into  Eq.(\ref{1}) we find after 
integration by parts of terms ${\bf nv_{1,2}}~({\bf nv_{1,2}} \rightarrow 1)$
the general expression for photon emission probability (or probability 
of creation of longitudinally polarized positron) 
\begin{eqnarray}
\hspace{-10mm}&& dw_{\xi} = \sigma \frac{\alpha m^2}{8 \pi^2}
\frac{d\Gamma}{\varepsilon\varepsilon'}
\int \frac{dN}{N} \int e^{i\sigma A}
\left[\varphi_1(\xi)-\frac{\sigma}{4}\varphi_2(\xi)\gamma^2\left(\textbf{v}_1-
\textbf{v}_2\right)^2\right]dt_1dt_2,
\nonumber \\
\hspace{-10mm}&& A=\frac{\omega \varepsilon}{2\varepsilon'}\int_{t_1}^{t_2}
\left[\frac{1}{\gamma^2}+(\textbf{n}-\textbf{v}(t))^2\right]dt,
\nonumber \\
\hspace{-10mm}&&\varphi_1(\xi)=1+\xi\frac{\omega}{\varepsilon}, \quad
\varphi_2(\xi)=(1+\xi)\frac{\varepsilon}{\varepsilon'}+
(1-\xi)\frac{\varepsilon'}{\varepsilon}.
\label{7.1}
\end{eqnarray}
where $\alpha=e^2=1/137$, the vector $\textbf{n}$ is defined 
in Eq.(\ref{1}), the helicity of emitted photon $\xi$
is defined in Eq.(\ref{5}), $\sigma=-1, d\Gamma=d^3k $ 
for radiation and
$\sigma=1,~d\Gamma=d^3p $ for pair creation and 
for pair creation one have to
put $\int dN/N=1$.

The circular polarization of radiation is defined by Stoke's parameter 
$\xi^{(2)}$:
\begin{equation}
\xi^{(2)}=\Lambda (\mbox{\boldmath$\zeta$}{\bf v}),\quad
\Lambda = \frac{dw_+ - dw_-}{dw_+ + dw_-},   
\label{7.1a}
\end{equation} 
where the quantity $(\mbox{\boldmath$\zeta$}{\bf v})$ defines the longitudinal
polarization of the initial electrons, $dw_+$ and $d_-$ is the probability
of photon emission for $\xi$=+1 and $\xi$=-1 correspondingly. In the limiting case
$\omega \ll \varepsilon$ one has $\varphi_2(\xi) \simeq 2(1+\xi\omega/\varepsilon)
=2\varphi_1(\xi)$. So the expression for the probability $dw_{\xi}$ contains the
dependence on $\xi$ as a common factor $\varphi_1(\xi)$ only. Substituting
in Eq.(\ref{7.1a}) one obtains the universal result independent of a particular
mechanism of radiation $\xi^{(2)}=\omega
(\mbox{\boldmath$\zeta$}{\bf v})/\varepsilon$.

The periodic crystal potential $U(\textbf{r})$ can be
presented as the Fourier series (see e.g.\cite{BKS}, Sec.8 )
\begin{equation}
U(\textbf{r})=\sum_{\textbf{q}}G(\textbf{q})e^{-i\textbf{q}\textbf{r}},
\label{8.1}
\end{equation} 
where $\textbf{q}=2\pi(n_1, n_2, n_3)/l;~l$ is the lattice constant.
The particle velocity can be presented in a form
$\textbf{v}(t) =\textbf{v}_0+\Delta\textbf{v}(t)$, where 
$\textbf{v}_0$ is the average velocity. If 
$\vartheta_0 \gg \vartheta_c$, we find $\Delta\textbf{v}(t)$
using the rectilinear trajectory approximation 
\begin{equation}
\Delta\textbf{v}(t)=-\frac{1}{\varepsilon}\sum 
\frac{G(\textbf{q})}{q_{\parallel}}\textbf{q}_{\perp}
\exp [-i(q_{\parallel}t+\textbf{q}\textbf{r})],
\label{9.1}
\end{equation} 
where $ q_{\parallel}=(\textbf{q}\textbf{v}_0),~
\textbf{q}_{\perp}=\textbf{q}-\textbf{v}_0(\textbf{q}\textbf{v}_0)$.
Substituting Eq.(\ref{9.1}) into Eq.(\ref{7.1}), performing the
integration  over ${\bf u}={\bf n}-{\bf v}_0 
(d^3k=\omega^2 d\omega d{\bf u},~d^3p=\varepsilon^2 d\varepsilon d{\bf u})$
and passing to the variables $t, \tau:~t_1=t-\tau,~t_2=t+\tau$, 
we obtain after simple calculations the general expression for       
the probability of photon emission (or probability of pair creation
) for polarized case valid for any angle of incidence 
$\vartheta_0$
\begin{eqnarray}
&& dw_{\xi}  
= \frac{i\alpha m^2 d\Gamma_1}{4 \pi}
  \int \frac{dN}{N} 
\int \frac{d\tau}{\tau+i\sigma 0}\Bigg[\varphi_1(\xi)+\sigma \varphi_2(\xi)
\nonumber \\
&& \times \left(\sum_{\textbf{q}} 
\frac{G(\textbf{q})}{m q_{\parallel}}\textbf{q}_{\perp}
\sin(q_{\parallel}\tau)e^{-i\textbf{q}\textbf{r}} \right)^2 \Bigg]
  e^{i\sigma A_2},
\label{10.1}
\end{eqnarray}
where 
\begin{eqnarray}
&& A_2=\frac{m^2\omega\tau}{\varepsilon\varepsilon'} \left[1+
\sum_{\textbf{q},\textbf{q}^{\prime} }\frac{G(\textbf{q})G(\textbf{q}')}
{m^2 q_{\parallel} q_{\parallel}'}
(\textbf{q}_{\perp}\textbf{q}_{\perp}')\Psi(q_{\parallel}, q_{\parallel}', \tau)
\exp [-i(\textbf{q}+\textbf{q}')\textbf{r}]\right] 
\nonumber \\
&& \Psi(q_{\parallel}, q_{\parallel}', \tau)=
\frac{\sin(q_{\parallel}+ q_{\parallel}')\tau}{(q_{\parallel}+ 
q_{\parallel}')\tau}-\frac{\sin(q_{\parallel}\tau)}{q_{\parallel}\tau}
\frac{\sin(q_{\parallel}'\tau)}{q_{\parallel}'\tau},
\label{11.1}
\end{eqnarray}
where $d\Gamma_1=\omega d\omega/\varepsilon^2
(\varepsilon d\varepsilon/\omega^2)$ for
radiation (pair creation).

\section{Radiation and pair creation for $\vartheta_0 \ll V_0/m$\\ 
(constant field limit)}

The behavior of probability $dw_{\xi}$ Eq.(\ref{10.1}) 
for various entry angles and energies is determined by 
the dependence of the phase $A_2$ on these parameters 
given Eq.(\ref{11.1}).  In the axial case for
$\vartheta_0 \ll V_0/m \equiv \vartheta_v$ the main contribution
to $A_2$ give vectors $\textbf{q}$ lying in the plane transverse 
to the axis \cite{BKS} and the problem becomes 
two-dimensional with the potential $U(\mbox{\boldmath$\varrho$})$.
Performing the calculation for axially symmetric 
$U=U(\mbox{\boldmath$\varrho$}^2)$ we obtain
\begin{eqnarray}
\hspace{-10mm}&& dw_{\xi}^F(\omega)=
\frac{\alpha m^2\omega d\Gamma_1}{2\sqrt{3}\pi}
\int_{0}^{x_0} \frac{dx}{x_0} \Bigg\{D R_0(\lambda)
 -\frac{1}{6}\left( \frac{m \vartheta_0}{V_0}\right)^2\Bigg[
\frac{xg''+2g'}{xg^3} R_1(\lambda)
\nonumber \\
\hspace{-10mm}&& -\frac{\lambda}{20g^4x^2}\left(2x^2g^{\prime 2}
+g^2+14gg^{\prime}x
+6x^2gg^{\prime \prime}\right)R_2(\lambda)\Bigg] \Bigg\}, 
\label{7.2}
\end{eqnarray}
where 
\begin{eqnarray}
\hspace{-8mm}&& R_0(\lambda)=\varphi_{2}(\xi)K_{2/3}(\lambda)+
\varphi_{1}(\xi)\int_{\lambda}^{\infty} K_{1/3}(y)dy,~
R_1(\lambda)=
\varphi_{2}(\xi)\Bigg[(K_{2/3}(\lambda) 
\nonumber \\
\hspace{-8mm}&&-\frac{2}{3\lambda}
K_{1/3}(\lambda)\Bigg],~  R_2(\lambda)=
\varphi_{1}(\xi)\left( K_{1/3}(\lambda) 
-\frac{4}{3\lambda}K_{2/3}(\lambda)\right)
\nonumber \\
\hspace{-8mm}&&-\varphi_{2}(\xi)\left(\frac{4}{\lambda}K_{2/3}(\lambda) 
-\left(1+\frac{16}{9\lambda^2} \right)K_{1/3}(\lambda) \right),
\label{6.2}
\end{eqnarray}  
here $D=\int dN/N (D=1)$ for the radiation (pair creation),
$K_{\nu}(\lambda)$ is the modified Bessel function (McDonald's function),
we have adopted a new variable $x=\mbox{\boldmath$\varrho$}^2/a_s^2,
~x \leq x_0,~x_0^{-1}=\pi a_s^2dn_a=\pi a_s^2/s,~a_s$ 
is the effective screening radius
of the potential of the string, $n_a$ is the density of atoms in a crystal,
$d$ is the average distance between atoms of a chain forming the axis.
The term in Eq.(\ref{7.2}) with $R_0(\lambda)$
represent the spectral probability in the constant field limit.
The other terms are the correction proportional $\vartheta_0^2$
arising due to nongomogeneity of field in crystal.
The notation $U^{\prime}(x)=V_0g(x)$ is used in Eq.(\ref{7.2}) and
\begin{equation}
\lambda=\frac{m^3a_s\omega}{3\varepsilon\varepsilon'V_0g(x)\sqrt{x}}=
\frac{u}{3\chi_s g(x)\sqrt{x}}=
\frac{2\omega^2}{3\varepsilon\varepsilon'}\frac{\sqrt{\eta}}{\kappa_s \psi(x)},
~\psi(x)=2\sqrt{\eta x} g(x),
\label{9.2}
\end{equation} 
here $\kappa_s=V_0\omega/(m^3 a_s),~\chi_s=V_0\varepsilon/(m^3 a_s)$ are
a typical values of corresponding parameters in crystal.
For specific calculation we use the following for the potential of axis:
\begin{equation}
U(x)=V_0\left[\ln\left(1+\frac{1}{x+\eta} \right)- 
\ln\left(1+\frac{1}{x_0+\eta} \right) \right]. 
\label{10.2}
\end{equation} 
For estimates one can put $V_0 \simeq Ze^2/d,~\eta \simeq 2u_1^2/a_s^2$,
where $Z$ is the charge of the nucleus, 
$u_1$ is the amplitude of thermal vibrations,
but actually the parameter of potential were determined by means of a fitting
procedure using the potential Eq.(\ref{8.1}) (table of parameters for 
different crystals is given in Sec.9 of \cite{BKS}).
For this potential
\begin{equation}
g(x)=\frac{1}{x+\eta} - \frac{1}{x+\eta +1} = \frac{1}{(x+\eta)(x+\eta +1)}
\label{11.2}
\end{equation} 

\begin{figure}[ht]
\centering
\includegraphics[width=14cm]{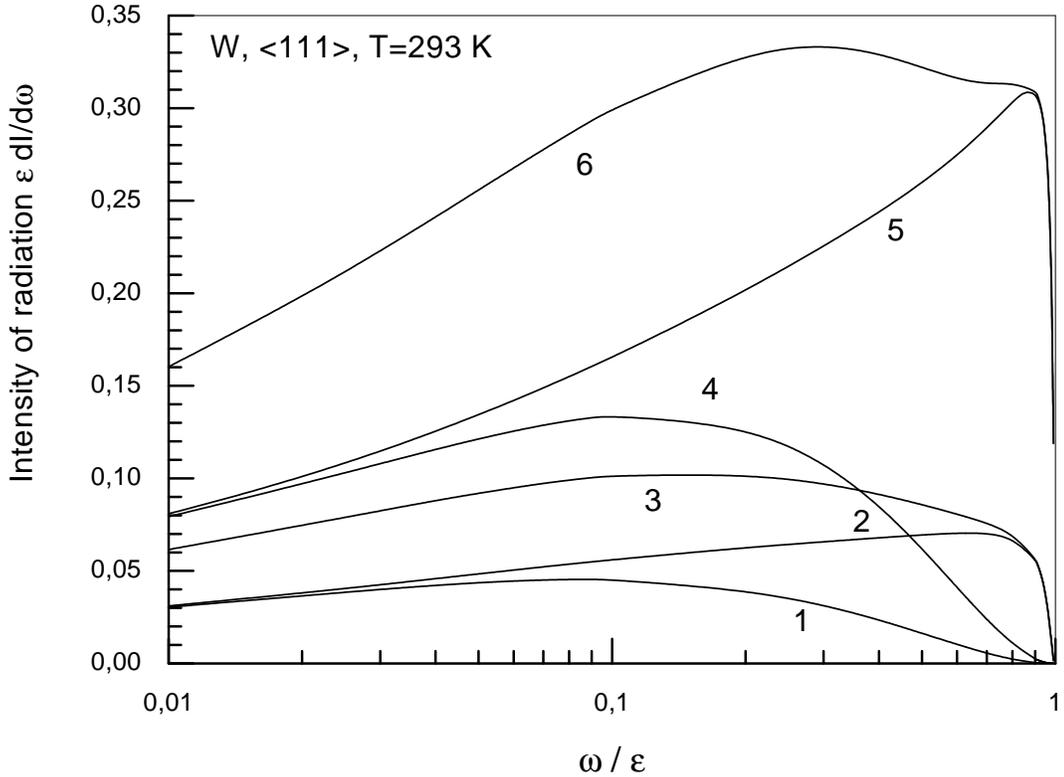}
\caption{Spectral intensity of radiation 
in units $\alpha m^2$ vs $\omega/\varepsilon$. 
The curves 1, 4 are for $\xi=-1$, the curves 2, 5 are for $\xi=1$, 
the curves 3, 6 are the sum of previous contributions 
(the probability for unpolarized particles).
The curves 1, 2, 3 are for the initial electron energy $\varepsilon$=250~GeV 
and the curves 4, 5, 6 are for the initial electron energy $\varepsilon$=1~TeV.}
\label{fig1}
\end{figure}

The result of calculation of spectral intensity 
$(dI_{\xi}=\omega dw_{\xi}),~
\displaystyle{\varepsilon\frac{dI_{\xi}^F(\omega)}{d\omega}}$ 
in tungsten crystal, 
axis $<111>$, $T=293~K$) is given in Fig.\ref{fig1}.

\begin{figure}[ht]
\centering
\includegraphics[width=14cm]{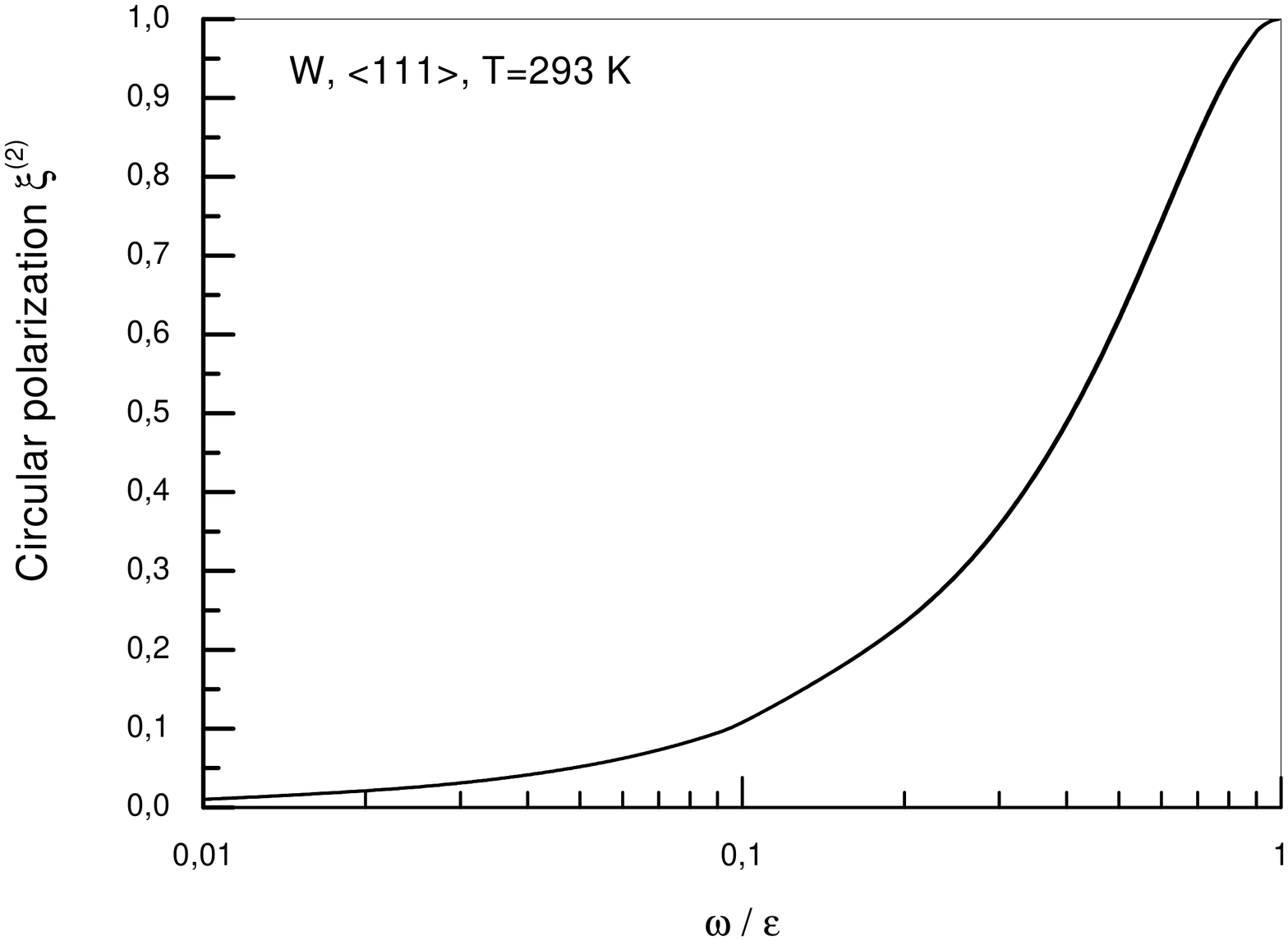}
\caption{The circular polarization $\xi^{(2)}$ 
of radiation (for ($\mbox{\boldmath$\zeta$}{\bf v}$)=1)
vs $\omega/\varepsilon$ for the tungsten crystal, 
axis $<111>$, $T=293~K$. The curve is valid for both energies:
$\varepsilon$=250~GeV and $\varepsilon$=1~TeV.}
\label{fig3}
\end{figure}

In Fig.\ref{fig3} the circular polarization $\xi^{(2)}$ of radiation is plotted 
versus $\omega/\varepsilon$ for the same crystal. This curve is true
for both energies:$\varepsilon=$250~GeV and  $\varepsilon=$1~TeV. 
Actually this means that it is valid for any energy in high-energy region.
At $\omega/\varepsilon$=0.8 one has $\xi^{(2)}$=0.94 and at 
$\omega/\varepsilon$=0.9 one has $\xi^{(2)}$=0.99.

\begin{figure}[ht]
\centering
\includegraphics[width=14cm]{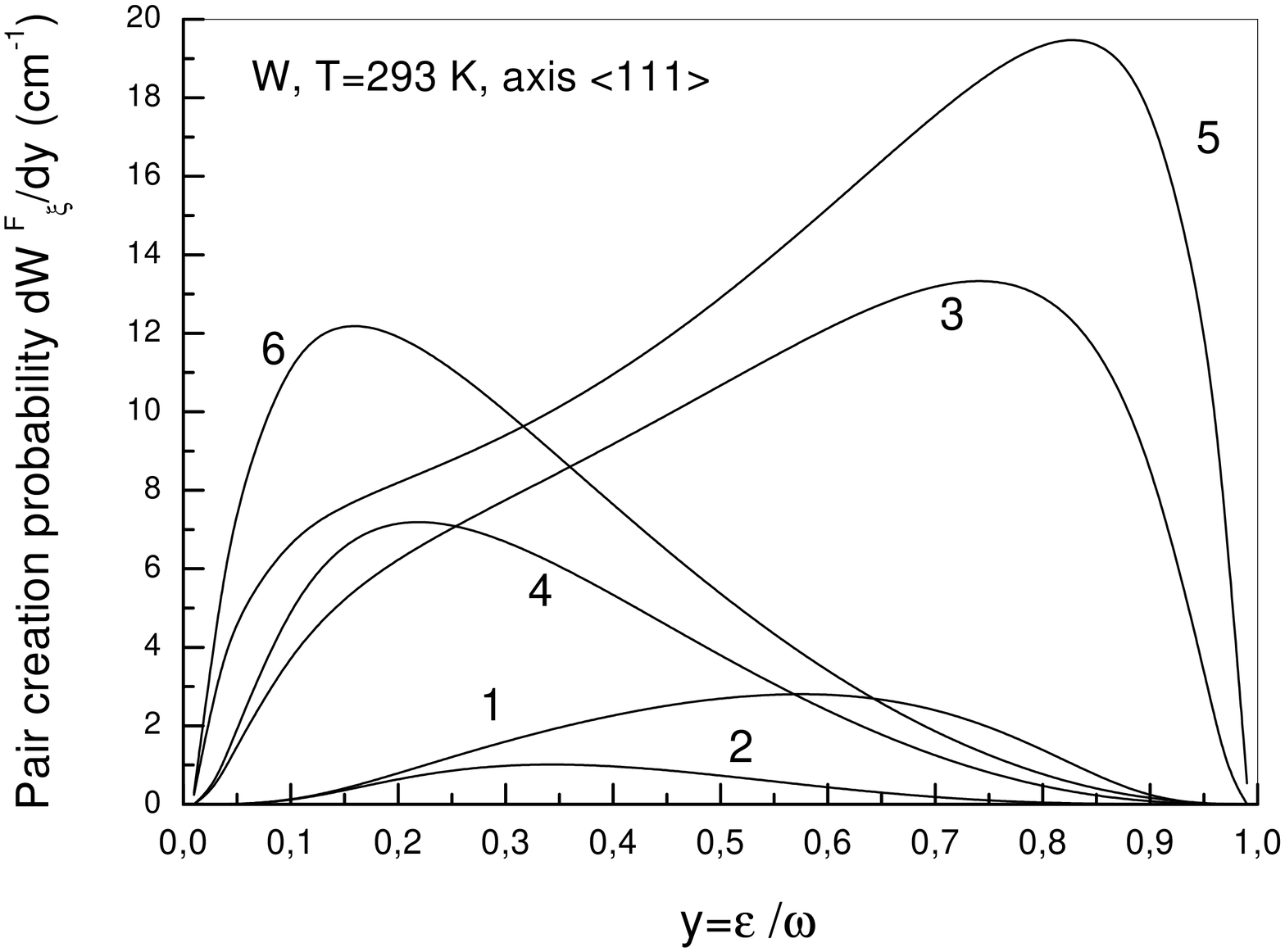}
\caption{The spectral probability of pair creation $dw_{\xi}^F/dy$,
the curves 1 and 2 are for energy $\varepsilon$=22~GeV,
the curves 3 and 4 are for energy $\varepsilon$=100~GeV,
the curves 5 and 6 are for energy $\varepsilon$=250~GeV.
The curves 1, 3 and 5 are for $\xi=1$, and
the curves 2, 4 and 6 are for $\xi=-1$.}
\label{fig2}
\end{figure}

Now we pass to a pair creation by a photon.
In the limit $\kappa_s \ll 1$ one can substitute 
the asymptotic of functions  
$K_{\nu}(\lambda)$ at $\lambda \gg 1$  in the term with $R_0(\lambda)$ 
in Eq.(\ref{7.2}). After this one obtains using the Laplace method in 
integration over the coordinate $x$
\begin{equation}
\frac{dW^F_{\xi}}{dy}=
\frac{\sqrt{3}\alpha V_0\left(y(1+\xi) 
+\left(1- y\right)^2 \right)}{2mx_0a_s}
\left[\frac{\psi^3(x_m)}{4\eta|\psi^{\prime\prime}(x_m)|} \right]^{1/2} 
e^{-\frac{2}{3y(1-y)\kappa_m}},
\label{13.2}
\end{equation}  
where $\kappa_m=\kappa_s \psi(x_m)/\sqrt{\eta}$. 
So, at low energies the pair creation probability is suppressed. 
With photon energy increase the probability increases also and
at some energy $\omega \simeq\omega_t$ it becomes equal to the standard 
Bethe-Maximon $W_{BM}$
probability in the considered medium, e.g. for W, T=293 K,
axis $<111>~\omega_t=22$~GeV. For high energies the probability of 
pair creation may be much higher than $W_{BM}$ (see \cite{BKS}).

This is seen in Fig.\ref{fig2} where the spectral 
probability of pair creation
$dw_{\xi}/dy$ in tungsten, T=293 K, axis $<111>$ is given. Near the end of
spectrum the process with $\xi=1$ dominates. The
sum of curves at the indicated energy gives unpolarized case.
For the integral (over $y$) probability the longitudinal polarization
of positron is $\zeta=2/3$.

\section{Modified theory of coherent bremsstrahlung and pair creation}

The estimates of double sum in the phase $A_2$ made at the beginning of previous 
section: $\sim (\vartheta_v/\vartheta_0)^2\Psi$ remain valid also for 
$\vartheta_0 \geq \vartheta_v$, except that now the factor in the double sum is
$(\vartheta_v/\vartheta_0)^2 \leq 1$, so that the values 
$|q_{\parallel}\tau| \sim 1$ contribute. We consider first the limiting case
$\vartheta_0 \gg \vartheta_v$, then this factor is small and $\exp(-iA_2)$
can be expanded accordingly.
After integration over coordinate $\textbf{r}$  we obtain
\begin{eqnarray}
\hspace{-10mm}&& dw_{\xi}^{coh}=\frac{\alpha  d\Gamma_1}{8}
\sum_{\textbf{q}} |G(\textbf{q})|^2 \frac{\textbf{q}_{\perp}^2}{q_{\parallel}^2}
\left[\varphi_2(\xi)+ \sigma \varphi_1(\xi)\frac{2m^2\omega}
{\varepsilon\varepsilon'q_{\parallel}^2}\left(|q_{\parallel}|- 
\frac{m^2\omega}{2\varepsilon\varepsilon'}\right)  \right] 
\nonumber \\
\hspace{-10mm}&& \times \vartheta\left(|q_{\parallel}|- 
\frac{m^2\omega}{2\varepsilon\varepsilon'}\right). 
\label{2.3}
\end{eqnarray} 
For unpolarized electrons (the sum of cotributions with $\xi=1$ and $\xi=-1$) 
Eq.(\ref{2.3}) coincides with the result of standard
theory of coherent bremsstrahlung (CBS), or coherent pair creation 
see e.g. \cite{TM}. 

In the case $\chi_s \gg 1$ ($\chi_s$ is defined in Eq.(\ref{9.2})), one can
obtain more general expression  for  the spectral
probability:
\begin{eqnarray}
\hspace{-10mm}&& dw_{\xi}^{mcoh}=\frac{\alpha  d\Gamma_1}{8}
\sum_{\textbf{q}} |G(\textbf{q})|^2 \frac{\textbf{q}_{\perp}^2}{q_{\parallel}^2}
\left[\varphi_2(\xi)+ \sigma \varphi_1(\xi)\frac{2m^2\omega}
{\varepsilon\varepsilon'q_{\parallel}^2}\left(|q_{\parallel}|- 
\frac{m_{\ast}^2\omega}{2\varepsilon\varepsilon'}\right)  \right] 
\nonumber \\
\hspace{-10mm}&& \times \vartheta\left(|q_{\parallel}|- 
\frac{m_{\ast}^2\omega}{2\varepsilon\varepsilon'}\right),\quad
m_{\ast}^2=m^2(1+\frac{\rho}{2}),\quad \frac{\rho}{2}=
\sum_{\textbf{q},q_{\parallel}\neq 0 }\frac{|G(\textbf{q})|^2}
{m^2 q_{\parallel}^2}
\textbf{q}_{\perp}^2
\label{2.4}
\end{eqnarray} 

The spectral probabilities Eqs.(\ref{2.3}) and (\ref{2.4}) can be
much higher than the Bethe-Maximon bremsstrahlung probability $W_{BM}$ for 
small angles of incidence $\vartheta_0$ with respect to selected axis.
For the case $\vartheta_0 \ll 1$ the quantity $q_{\parallel}$
can be represented as
\begin{equation}
q_{\parallel} \simeq \frac{2\pi}{d}n+\textbf{q}_{\perp}\textbf{v}_{0.\perp}  
\label{5.3}
\end{equation} 

In the extreme limit,
when the parameter $\lambda \equiv 2\varepsilon|q_{\parallel}|_{min}/m^2
\sim \varepsilon \vartheta_0/m^2 a_s \gg 1$, the maximum 
of probability of coherent bremsstrahlung is attained at such values of
$\vartheta_0$ where the standard theory of coherent bremsstrahlung
becomes invalid. Bearing in mind that if $\lambda \gg 1$  and 
$\vartheta_0 \sim V_0/m$, then $\chi_s \sim \lambda \gg 1$,
we can conveniently use a modified theory of coherent bremsstrahlung.

The direction of transverse components of particle's velocity in 
Eq.(\ref{5.3}) can be selected in a such way, that
the spectral probability described by Eq.(\ref{2.4}) has a sharp maximum
near the end of spectrum at 
$\omega = 2\varepsilon\lambda(2+2\lambda+\varrho)^{-1} \simeq \varepsilon$
with relatively small (in terms of $\lambda^{-1}$) width 
$\Delta \omega \sim \varepsilon(1+\varrho/2)/\lambda=
m^2(1+\varrho/2)/2 |q_{\parallel}|_{min}$:
\begin{equation}
\left(dw_{\xi}\right)_{max}=
\frac{\alpha \varepsilon d\Gamma_1 \varrho |q_{\parallel}|_{min}}{4(2+\varrho)}
\left(1+\xi + \frac{1-\xi}{(1+u_m)^2}\right),\quad 
u_m=\frac{2\lambda}{2+\varrho}.
\label{8.3}
\end{equation} 
It is seen that in the maximum of spectral distribution the radiation 
probability with opposite helicity ($\xi=-1$) is suppressed as 
$1/(1+u_m)^2$. At $u > u_m$ one have to take into account the next
harmonics of particle acceleration. In this region of spectrum 
the suppression of  radiation probability with opposite helicity is more strong,
so the emitted photons have nearly complete circular polarization.

Comparing Eq.(\ref{8.3}) with $W_{BM}$
for $\varepsilon' \ll \varepsilon$ we find that for the same circular 
polarization ($\xi^{(2)} \simeq (\mbox{\boldmath$\zeta$}{\bf v})$) in the 
particular case $\varrho = 1$ the magnitude of spectral probability in 
Eq.(\ref{8.3}) is about $\chi_s(\varepsilon_e)$ times larger than
$W_{BM}$. For tungsten $\chi_s(\varepsilon_e)=78$. 
From above analysis follows that under mentioned conditions the
considered mechanism of emission of photons with circular polarization
is especially effective because there is gain both in monochromaticity
of radiation and total yield of polarized photons near 
hard end of spectrum.

\section{Conclusions}

At high energy the radiation from 
longitudinally polarized electrons in oriented crystals is circularly
polarized and $\xi^{(2)} \rightarrow 1$ near the end of spectrum.
This is true in magnetic bremsstrahlung limit $\vartheta_0 \ll V_0/m$
as well as in coherent bremsstrahlung region $\vartheta_0 > V_0/m$. 
This is particular case of helicity transfer.

In crossing channel: production of 
electron-positron pair with longitudinally polarized particles
by the circularly polarized photon in an oriented crystal the same phenomenon
of helicity transfer takes place in the case when the final particle 
takes away nearly all energy of the photon.

So, the oriented crystal is a very effective device for helicity transfer 
from an electron to photon and back from a photon to electron or positron.
Near the end of spectrum this is nearly 100\% effect.

\vspace{0.25 cm}
{\bf Acknowledgements}
\vspace{0.25 cm}
We would like to thank the Russian Foundation for Basic
Research for support in part this research by Grant 
03-02-16154.


\end{document}